\documentclass[12pt,letter]{article}
\usepackage{authblk}
\usepackage{graphicx}
\usepackage{amsmath}
\usepackage{amssymb}
\usepackage[left=1in,right=1in,top=1in,bottom=1in]{geometry}
\usepackage{setspace}
\usepackage{float}
\usepackage[font={small}]{caption} 
\usepackage[labelfont=bf]{caption}
\usepackage[superscript,biblabel]{cite}
\usepackage{bm}
\usepackage[T1]{fontenc}
\usepackage{microtype}
\usepackage{subfigure}
\usepackage{chemformula}
\usepackage[version=4]{mhchem}
\usepackage{textcomp}
\usepackage{color}
\usepackage{mathtools}
\usepackage{bm}
\usepackage{gensymb}
\usepackage{esvect}
\usepackage{booktabs}
\usepackage{multirow}
\usepackage{indentfirst}
\usepackage{mathrsfs}
\usepackage{float}
\usepackage{hyperref}
\usepackage{xcolor}
\usepackage{booktabs}
\usepackage{amssymb} 
\usepackage{bm} 
\usepackage{tablefootnote}
\usepackage{enumitem}
\setlist{leftmargin=1.5em}
\newcommand*{\addFileDependency}[1]{
  \typeout{(#1)}
  \@addtofilelist{#1}
  \IfFileExists{#1}{}{\typeout{No file #1.}}
}
\makeatother

\begin{document}

\title{\textbf{Foundation Models for Atomistic Simulation of Chemistry and Materials }}

\author{Eric C.-Y. Yuan$^{1,6}$, Yunsheng Liu$^{1}$, Junmin Chen$^{1}$, Peichen Zhong$^{5}$, Sanjeev Raja$^{4, 7}$, Tobias Kreiman$^{4}$, Santiago Vargas$^{6,8}$, Wenbin Xu$^{9}$, Martin Head-Gordon$^{1,6}$, and Chao Yang$^{7}$\\ 

Samuel M. Blau*$^{8}$, Bingqing Cheng*$^{1,6}$, Aditi Krishnapriyan*$^{3,4,7}$, Teresa Head-Gordon*$^{1-3,6}$}
\date{}
\maketitle
\begin{center}
\vspace{-10mm}
$^1$Kenneth S. Pitzer Theory Center and Department of Chemistry, $^2$Department of Bioengineering, $^3$Department of Chemical and Biomolecular Engineering, $^4$Electrical Engineering and Computer Science, $^5$Bakar Institute of Digital Materials for the Planet, University of California, Berkeley, CA, 94720 USA\\

$^6$Chemical Sciences Division, $^7$Applied Mathematics and Computational Research Division, $^8$Energy Technologies Area, $^9$National Energy Research Scientific Computing Center, Lawrence Berkeley National Laboratory, Berkeley, CA, 94720 USA

*corresponding authors: aditik1@berkeley.edu, bingqingcheng@berkeley.edu, smblau@lbl.gov, thg@berkeley.edu
\end{center}

\begin{abstract}
\noindent
Given the power of large language and large vision models, it is of profound and fundamental interest to ask if a foundational model based on data and parameter scaling laws and pre-training strategies is possible for learned simulations of chemistry and materials. The scaling of large and diverse datasets and highly expressive architectures for chemical and materials sciences should result in a foundation model that is more efficient and broadly transferable, robust to out-of-distribution challenges, and easily fine-tuned to a variety of downstream observables, when compared to specific training from scratch on targeted applications in atomistic simulation. In this Perspective we aim to cover the rapidly advancing field of machine learned interatomic potentials (MLIP), and to illustrate a path to create chemistry and materials MLIP foundation models at larger scale. 
\end{abstract}

\newpage
\section{Introduction}
\label{sec:intro}
\noindent
The fundamental principle underlying the description of atomistic chemistry and materials science is the Schrödinger equation, whereby a chemical system and its physical properties can be uniquely defined by Cartesian and spin coordinates of the component atoms and system net charge. But as famously stated by Dirac after the discovery of quantum mechanics "..the underlying physical laws necessary for the whole of chemistry are thus completely known, and the difficulty lies only in the fact that the exact application of these laws leads to equations much too complicated to be soluble." This has led to a series of well-controlled approximations, such as the Born-Oppenheimer separation of nuclei and electronic degrees of freedom, and the transformation of the Schrödinger eigenvalue equation into an algebraic framework, that has allowed for good model chemistries to be solved for systems beyond the hydrogen molecule.\cite{Pople1973} Just as important is the need to satisfy the laws of statistical mechanics that requires the high dimensional integration over coordinates of Boltzmann weighted configurations to predict observables, usually collected over time trajectories with the hope that ergodicity is satisfied. Even so, the most widely used approximate electronic structure methods based on Density Functional Theory (DFT)\cite{Mardirossian2017} combined with ab initio molecular dynamics (AIMD) remains inaccessible for long timescales and large length scales and system sizes that define many of the interesting areas of chemistry and materials science. 

How might we resolve this tension between the need for quantum mechanical accuracy with the need to satisfy statistical mechanical sampling to yield converged and correct observables? One recent possibility is to leverage the tools of machine learning and data science. Their success in other domains suggests that we could learn ``the foundations'' of molecular and materials chemistry from an abundance of data using appropriate architectures and training strategies. The concept of a foundation model (FM) has been established by the natural language processing and computer vision communities through the development of large language models (LLMs) and large vision models (LVMs).\cite{Hestness2017,Bommasani2021} These large parameter architectures are pre-trained in an unsupervised or self-supervised manner on enormous amounts of data, and can then be easily fine-tuned for accurate prediction on seemingly unrelated downstream tasks, revolutionizing the field of artificial intelligence. 

While LLMs and LVMs themselves hold great potential to impact chemistry and materials science\cite{Zheng2023,Boiko2023,Cavanagh2024,Bran2024,Sun2025}, in this perspective we are focused on whether FMs can directly capture the potential energy surface (PES) via machine learned interatomic potentials (MLIPs).\cite{Unke2021_MLFF, Wang2024_MLIP} The end goal of such a model is to enable large-scale atomistic simulation and analysis, including the ability to derive any observable of interest, for any molecular or periodic system, either with a small amount of system-specific fine-tuning\cite{Allen2024_metalearning} or without any fine-tuning at all. Importantly, it would supersede the painful acquisition of new labeled data and the hidden expense of often hundreds of rounds of retraining of the ML model used in active learning to develop a robust MLIP.\cite{Smith2018,Khalak2022,Kulichenko_Barros_2023,Guan2023}

To achieve such a goal, this perspective first provides more concrete definitions for what defines a FM, and how a FM for atomistic simulation should be distinguished from a ``universal potential'' or even a more restricted definition such as transfer learning. In particular, our criteria begins with the ability of a MLIP FM to exhibit scaling laws and pre-training and data strategies that have made LLMs and LVMs so powerful. With this context, we examine the capacity of current model architectures and identify limitations and opportunities that are relevant to their scalability to larger learnable parameters in order to become more expressive and foundational to increasing amounts of data. We also consider what type and amount of data of scale will be required for a generally applicable pre-trained FM, how that compares to the data that already exists, and where we expect fine-tuning data to be most important for chemistry and materials. In this perspective we focus primarily on FMs trained on energy and force data to address outstanding problems in chemistry and materials through molecular simulation. But it should be pointed out at this juncture that different modality data exists, such as images and text extraction for catalytic discovery\cite{nandy_jcatal_perspective_2025}, prompt engineering of LLMs for predicting synthetic pathways of small drug molecules\cite{Sun2025}, and use of video data such as that generated for liquid-phase TEM imaging\cite{Yao2020}. We further examine the importance of more advanced training strategies, including self-supervised and unsupervised pre-training as well as model distillation, and highlight that new methods for physical infusion need to be relevant and feasible when training at scale on advanced hardware. Finally, we consider how the performance of MLIPs is currently evaluated, and to suggest new benchmarks for demonstrating scaling laws and the ability to be broadly fine-tuned for a wide range of downstream applications, which is particularly important given the breadth of desired observables in chemistry and materials. 

\section{Characteristics of Foundation Models}
\label{sec:def}
\noindent
The term ``foundation model'' has been loosely defined to mean any large-scale model pre-trained on large amounts of diverse data to capture a broad range of complex patterns\cite{Bommasani2021}, as represented by popular LLMs such as GPT\cite{Brown2020} and Llama\cite{Touvron2023}. One of their defining characteristics is that they often obey heuristic scaling laws\cite{Hestness2017,Kaplan2020,Hoffmann2022,Bahri2024}, which refers to how their performance improves as a function of increasing model size, training data, and computational resources. Increasing model capacity improves expressivity and performance when coupled with improved algorithms for neural network operations, such as attention mechanisms most commonly used in highly scalable Transformer \cite{vaswani2017attention} architectures. Data scaling for pre-training refers to increasingly large and diverse datasets of typically unlabeled data to learn general features and patterns, covering various domains and contexts to ensure broad knowledge acquisition. Compute scaling permits parameter and data scaling through optimal use of GPU and CPU hardware, exploiting single and mixed precision computation when appropriate, and taking advantage of parallelization using distributed computing. Exhibiting and satisfying these scaling laws has been critical for FMs in achieving strong results across various applications and domains, ranging from natural language processing, computer vision, climate modeling, and robotics.\cite{Brohan2022,Brohan2023,Kim2024,Octo2024} 

Coupled with the scaling laws is a pre-training strategy that ultimately yields a high efficiency model that performs well across a series of benchmark data sets that exist for fine-tuning for down-stream tasks. Pre-training objectives, particularly in language, often involve self-supervised learning tasks such as predicting missing words in sentences or predicting the next word in a sequence. During the earlier stages of the training process, more compute is used while data quality is typically lower and human supervision is minimal. The resulting pre-trained scaling models, data, and compute can lead to emergent capabilities, where a FM becomes capable of solving a task that appeared not possible at smaller scales \cite{Wei2022}. 
Although it is still disputed how and why capabilities are unlocked at scale \cite{Schaeffer2023}, these laws provide a quantitative framework for understanding and predicting the behavior of chemical FMs as they are scaled up, and offer insights into how different factors (architecture, data, pre-training) contribute to model performance improvements.

One important clarification that we hope to address in this perspective is to distinguish the term ``foundation model" from other related terms such as ``transferable''  or ``universal'', or large atomic models (LAMs)\cite{Zhang2024}, particularly in the context of MLIPs. The popular MLIP MACE-MP-0\cite{Batatia2023} is trained on data from the Materials Project\cite{Jain2013, Deng2023} generated at the PBE level of DFT. The resulting ML model exhibits good accuracy in energy and force predictions against PBE on other crystalline unit cells and exhibits excellent stability in the numerical solution to Hamilton's equation of motion across related chemical and materials systems. This has opened up a whole new range of applications, including free energy calculations, simulating spectroscopic observables, and explorations of phase diagrams. However, we would stipulate that MACE-MP-0 is not a foundation model, but an excellent universal PBE potential, given that it is trained with supervision to do a very specific task: predict energy and force labels for a single chemical domain at one level of theory. Another key limitation is that MACE-MP-0\cite{Batatia2023}, like many universal MLIPs, does not attempt to distinguish between systems with different total charge and spin, whose energy and forces would be different although the underlying input structure is the same. 

We offer that to be a true FM preferably all of the following features should be demonstrated. First is that the zero-shot or fine-tuned MLIP FM should show performance across a broad range of down-stream tasks that is superior to task-specific models trained from-scratch \cite{Zhang2024,Kaur2025}. Second, MLIP FMs would show compliance with heuristic scaling laws in regards how performance improves as model parameters and training data increases with increasing compute resources. Finally, a demonstration that a large-scale FM MLIP would have emergent capabilities, for example predicting higher quality CCSD(T) data or magnetic field-dependent properties, properties that are very different than the original underlying Cartesian geometries and DFT energy training data.

\section{Foundation Models for Atomistic Simulations}
\label{sec:chem}
\noindent
\subsection{Machine Learning Architectures for Chemistry and Materials }
\label{subsec:model}
\noindent
Over the past two decades, MLIP architectures have evolved significantly, integrating the inductive biases of chemistry while incorporating advancements from other areas of machine learning. We provide a general survey of these developments, as well as an outlook on recent architectures which increasingly exploit scale as a driving factor in performance. The architectures of MLIPs are posited to be distinct from those developed in computer science for tasks like natural language processing, the reason being given that chemistry is governed by physical laws and constraints that are more inviolable than the statistical nature of probability distributions in language. Hence most MLIP architectures are typically "physics-informed", while the exact manner in how to realize the physical constraints has gone through a few waves of evolution.

The early Behler-Parrinello neural network potentials\cite{Behler2007} used hand-crafted two-body and three-body descriptors for describing atomic environments. Later, DeePMD models\cite{Zhang2018} automated the optimization and discovery of such descriptors, reducing reliance on manual feature engineering. The atomic cluster expansion (ACE) method\cite{Drautz2019} introduced a unified and generalizable framework for constructing atom-centered descriptors using systematically improvable body-ordered terms. Graph neural networks (GNNs) then broadened the landscape, allowing for flexible and expressive architectures that learn features by iteratively exchanging information between neighboring atoms, thus capturing many-body effects\cite{Gilmer2017,Lubbers2018,Batzner2021,Haghighatlari2022}. More recently, GNN models have incorporated different forms of graph attention mechanisms.~\cite{Liao2022,Qu2024}. These models can be viewed as a more general way of performing message passing between atoms. 

Despite the rapid emergence of diverse architectures, most MLIPs incorporate similar inductive biases. The first is the nearsightedness assumption. This postulates that energy and forces experienced by a central atom are primarily influenced by its neighboring atoms within a finite cutoff radius. Nearsightedness readily ensures the key model chemistry requirement of size-consistency\cite{Pople1973} is satisfied. Message passing schemes relax this assumption to some extent by enabling atoms to aggregate information from progressively distant neighbors through multiple iterations, although intermediate neighbors still contribute the most. Second are the physical constraints such as energy conservation, smoothness, and invariances (translational, rotational, and permutational). These constraints are incorporated either explicitly within the model architecture (e.g., through gradient-based forces, symmetrized input features or invariance-preserving transformations) or implicitly via the loss function and training data. The third is to describe interatomic interactions via body-order terms. Two-body (pairwise) terms dominate, while three-body terms are also critical. Higher-body terms exhibit diminishing returns due to increased computational cost and reduced relative contributions. Frameworks like the Atomic Cluster Expansion (ACE)\cite{Drautz2019} provide systematic control over the inclusion of higher-body terms, with user-specified truncation. In GNNs, adding a message-passing layer effectively increases the body order by one to yield models of different ranks. 

Moreover, the seemingly diverse existing architectures share the same mathematical foundation. GNNs are all based on the same representations where atoms serve as nodes and their distances within the cutoff radius are edges. Additionally, almost all atomic and structural representations for materials and molecules are fundamentally related and can be understood within the unifying framework of atomic density\cite{Liu2019,Musil2021} or, equivalently, ACE\cite{Drautz2019}. Moreover, these representations can also be viewed as specific graph features, encoding geometric and chemical information of the nodes. This duality highlights a deeper connection: most existing MLIPs can be understood as different variations of the same underlying framework. 

The unifying framework of the current MLIP architectures and the common approach to enforce physical constraints are to be contrasted with the well-known ``bitter lesson'' that argues that general methods that leverage computation and data are more effective in the long term over approaches that try to build knowledge into the ML pipeline\cite{Sutton2019}. The main limitations of constraints are the challenges associated with the training dynamics and resulting poor loss landscapes, as well as difficulty with parallelization and distributed training on modern GPU hardware~\cite{marquez2017imposing, dosovitskiy2021an, gruver2023the}. Indeed, we are now seeing that scalable architecture methods that emphasize expressive capacity can both outperform models with built-in constraints and effectively learn these constraints directly from the data, while being significantly more compute-efficient.\cite{Qu2024,Octo2024, Kim2024, grattafiori2024llama3herdmodels, yu2022pointbertpretraining3dpoint, Abramson2024_alphafold3}. This offers the first evidence of the important bitter lesson learned from other fields of ML that also have constraints, in that fewer inductive biases do seem to eventually work better. It further suggests that satisfying data and parameter scaling for chemistry and materials will be more effective compared to models that build constraints into the model.

\begin{figure}[H]
 \centering
 \includegraphics[width=0.95\textwidth]{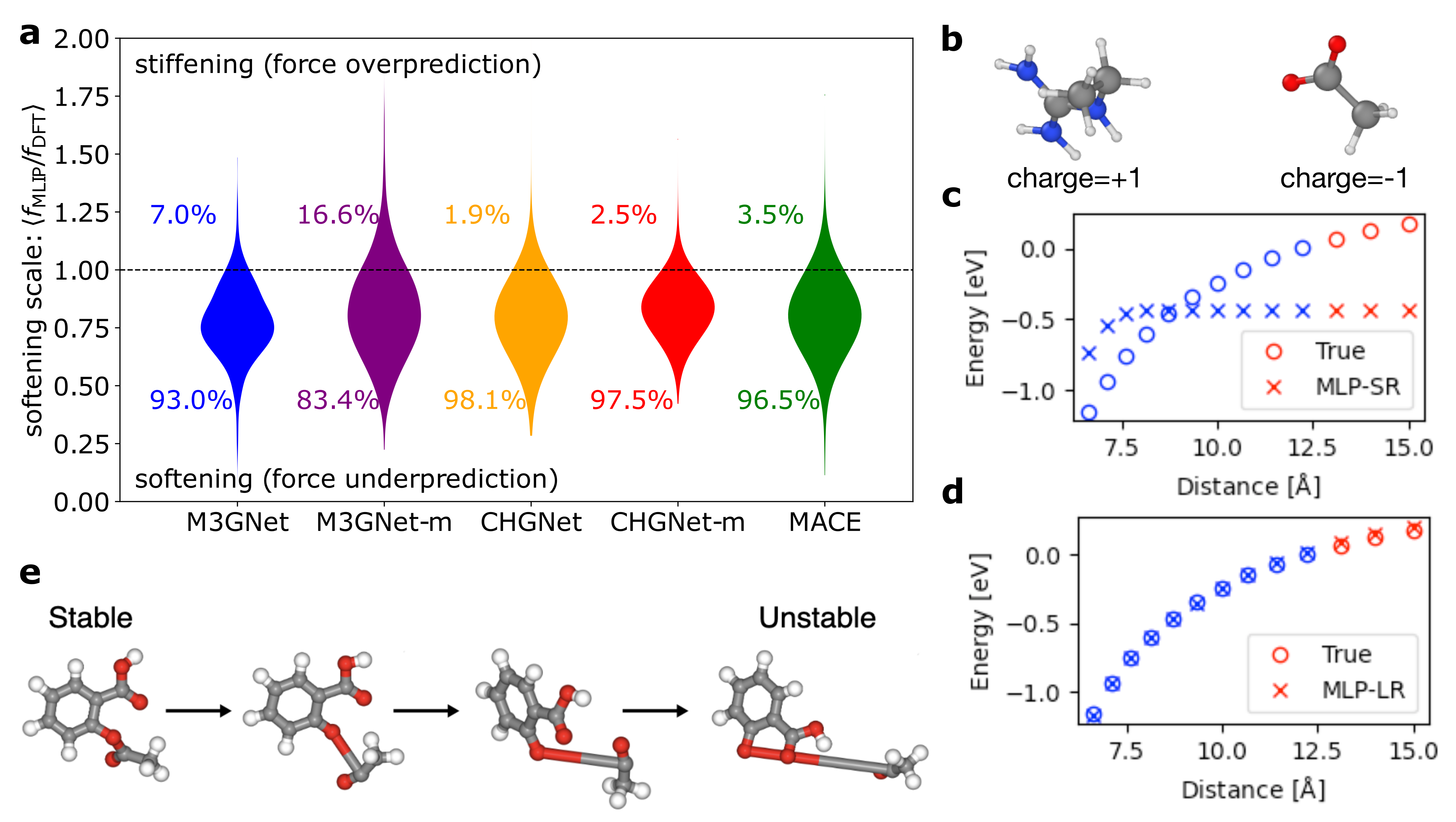}
\caption{(a) Distribution of softening scales ($f_{\text{MLIP}} / f_{\text{DFT}}$) sampled from 1,000 high-energy configurations using the structures from the WBM dataset.\cite{WBM_2021} Reproduced with permission from Ref. \cite{Deng2025}.(b) An example of two charged molecules that are far apart and do not interact due to (c) short-range implicit bias from GNNs and message passing, and (d) a new ML model that incorporates long-ranged information.~\cite{Cheng2024}. Blue points are for training and red points are for testing. Adapted with permission from Ref. \cite{Cheng2024} (e) Depiction of instability in molecular simulations with MLIPs, which can occur even at very small time discretizations~\cite{raja2025stabilityaware}. Adapted with permission from Ref. \cite{raja2025stabilityaware}}
 \label{fig:longrange}
\end{figure}

One of the most well-studied physical constraints is rotational equivariance.\cite{Batzner2021,Haghighatlari2022,Liao2023} It is worth noting that rotational equivariance is an easy property to satisfy, and cheap data augmentation (rotational transformations of the existing data) is more than adequate in nearly all cases \cite{Liu2019,Neumann2024,Qu2024}. It is a more open question whether more complicated physical laws can be learned from large data or still require some inductive biases. An example would be whether the symplectic structure of Hamilton's equation of motion can be respected, i.e. to drive stable molecular dynamics (MD) simulations. Stability in MD requires two key conditions: the first is the absence of pathological behaviors, such as molecules "exploding" due to data insufficiencies and numerical instabilities, even at very small time discretizations (Fig.\ref{fig:longrange}e). Second, is the ability to perform MD simulations without introducing non-conservative and energy drift artifacts. Recent studies have observed that certain MLIPs that use directly predicted non-conservative forces rather than obeying the energy conservation law can lead to issues such as artificial heat generation in NVE ensembles or severely modified dynamics in NVT ensembles~\cite{Bigi2024}. These behaviors undermine the reliability of such non-conservative MLIPs for long-timescale dynamics and their integration into predictive downstream workflows. We also note that energy conservation is a constraint that can be agnostic to the architecture itself, as the standard way it is enforced is by taking gradients of the predicted potential energy in the loss function \cite{Haghighatlari2022}, an expense that is avoided in recent direct force models\cite{Liao2023,Neumann2024}.

There are possible issues of the current MLIP architectures that should be examined more closely. The first is transferability outside the pre-training data. While recently developed universal MLIPs demonstrate reasonable performance in predicting materials stability on element-wise out-of-distribution datasets\cite{WBM_2021, Riebesell2023}, these evaluations primarily focus on local energy minimum searches through structure relaxation of crystalline materials, sharing similar prototypes in the Materials Project database. A systematic softening (underestimation) of the ML-predicted PESs and interatomic forces\cite{kreiman2025understandingmitigatingdistributionshifts} compared to DFT ground truth was observed in a recent benchmark study as seen in Figure \ref{fig:longrange}a. \cite{Deng2025} The underestimation of the PES is particularly significant for high-energy atomic configurations, which limits its application to molecular dynamics simulations and rare-event samplings. Recent large data efforts\cite{Zhang2024,BarrosoLuque2024} and open-sourced universal MLIPs\cite{Yang2024, mazitov2025pet} that include more off-equilibrium atomic configurations show promise in addressing these challenges as discussed in Section \ref{subsec:data}. However, direct transferability of universal MLIPs for specific downstream molecular or materials modeling remains uncertain, as they are not yet demonstrably better than models built directly for the new system or property from scratch. We return to this issue in Section \ref{subsec:eval}.

Another limitation is the lack of incorporation of long-range interactions. Although short-range potentials may be sufficient to describe most properties of homogeneous bulk systems \cite{Yue2021}, they may fail for interfaces \cite{Niblett2021},
dielectric response \cite{Rodgers2008,Cox2020}, and dilute ionic solutions with Debye-H\"{u}ckel screening. For example, consider two charged molecules that are separated by a distance greater than the cutoff radius used in most MLIPs (Fig.\ref{fig:longrange}b). These models, which focus on short-range interactions, will fail to accurately describe the electrostatic forces between the molecules (Fig.\ref{fig:longrange}c). This issue cannot be resolved simply by using message passing, as the molecules exist on separate graphs and do not interact within the framework of short-range biases inherent to message-passing neural networks. On the other hand, MLIPs with built-in long-range corrections~\cite{Grisafi2019,Cheng2024,kim2024learning} can effectively solve such artifacts (Fig.\ref{fig:longrange}d).

There is also the concern of scalability: whether scaling laws will be exhibited as the community acquires larger and larger data sets and computational resources. Is the GNN architecture adequate, or is there a need to shift towards more scalable architectures \cite{DiGiovanni2023, Rusch2023}? Relatedly, computational inefficiency when applying MLIPs at scale \cite{Sriram2022} can limit the application to realistic materials and chemical problems. The current short-ranged MLIPs are already orders of magnitude more expensive than traditional force fields. Further incorporating long-range interactions may require more computational power, and while there is a willingness to pay this price for improved accuracy, the current MLIP frameworks are not fully optimized for such tasks. While GNNs provide a powerful framework for capturing atomic interactions, their computational inefficiency at scale necessitates innovative approaches that can maintain or even enhance their model performance. 

\subsection{Data Requirements, Pitfalls, and Challenges}
\label{subsec:data}
\noindent
Real-world image and language datasets often contain trillions of tokens, whereas simulated chemistry datasets for training MLIPs typically contain between 1 to 200 million molecule or material configurations with associated energy and force data. Thus, increases in dataset size remain critical for improving the pre-training of MLIPs if they are to ultimately serve as robust FMs. Table \ref{tab:data} summarizes molecular and materials datasets with over 1,000,000 energy- and force-labeled structures that have emerged over the last $\sim$5 years. The underlying labeling methods are DFT calculations with variable basis set sizes. We also include large-scale non-DFT chemical data such as Uni-Mol2\cite{Ji2024}, which is a curated subset of the Zinc20 molecular data set\cite{Irwin2020}, and more recently Zinc22\cite{Tingle2023} which is comprised of 4.5 billion molecules that also contains force field derived labels of their physical properties, including partial atomic charges, cLogP values, and solvation energies\cite{Tingle2023}. In all cases this variable quality in data is highly appropriate for pre-training where more value should be placed on broadly covering chemical space, with domain diversity i.e. molecular, interfacial, and solid-state systems, elemental diversity, diversity in both total charge and spin, molecular/structural diversity covering possible bonding and interaction motifs, and finally configurational diversity including both near-minima and far-from-minima configurations. 

Regarding elemental diversity, most plane-wave DFT datasets follow the standards set by the Materials Project\cite{Jain2013} and thus cover most of the periodic table up through plutonium. In contrast, molecular force field and DFT datasets are considerably more constrained in their elemental coverage, with most datasets only including C, H, N, and O and some coverage of S, F, Cl, and Br. The only molecular datasets of over 1 million data points with both energy and forces labels and which go beyond this set of elements are SPICE\cite{Eastman2023} (P, Li, Na, Mg, K, Ca, and I), AIMNet2 (P, I, Si, B, As, and Se), and more recently QCML (73 elements total)\cite{Ganscha2025} and OMol25 (83 elements)\cite{levine2025}. The only large-scale datasets with variable-charge species are SPICE(2)\cite{Eastman2023}, AIMNet2\cite{Anstine2024}, AIMNet-NSE\cite{Zubatyuk2021}, solvated protein fragments\cite{Unke2019}/GEMS\cite{Unke2024}, and most recently QCML\cite{Ganscha2025} and  OMol2025\cite{levine2025}. While AIMNet-NSE\cite{Zubatyuk2021} is a dataset with variable spin species, it carries no force labels, whereas QCML does include both labels for some limited spin species\cite{Ganscha2025} whereas variable spin is much better covered by OMol2025\cite{levine2025}.
Another consideration for chemical diversity is the variety of chemical interactions captured by the data. While most of the molecular datasets mentioned thus far focus on isolated molecules, SPICE\cite{Eastman2023} makes an effort to capture varied intermolecular interactions between molecules, including some metal ions, solvated protein fragments\cite{Unke2019}/GEMS\cite{Unke2024} includes protein fragments with multiple nearby solvent species, and ANI-2x\cite{Devereux2020} includes dimer scans and water clusters. Recognizing the value of intermolecular interactions, OMol25 also includes these data types.\cite{levine2025}

In order to create structural diversity, large-scale data-acquisition efforts use a collection of sampling techniques such as normal modes, ab initio and empirical molecular dynamics, minimization and relaxation, reaction path sampling, metadynamics, and active learning, as outlined in Table \ref{tab:data}. Structural diversity also encompasses the nature of the regions of the PES covered by the dataset. Having a broad range of forces between 0 eV/\AA\ and $\sim$10 eV/\AA\ is important to effectively learn chemically-critical components of the PES encountered during MD, geometry optimization, or reaction path and transition state optimization. To that end, Transition-1x is the only large-scale dataset which explicitly samples reaction paths and transition states -- critical areas for any truly foundational MLIP.\cite{Yuan2024,Wander2024}  Extremely high 

\begin{table}[H]
\centering
\footnotesize
\begin{tabular}{lllll}
\toprule
\toprule
\bf Molecular Dataset & \bf Size & \bf Calculation method(s) & \bf Sampling method(s) \\
\midrule
OMol25\cite{levine2025} & 100M & $\omega$B97M-D3/def2-TZVPD 
     & Molecular dynamics \\ 
 & & & Rattling \\ 
  & & & Geometry optimization \\ 
    & & & Reaction pathways \\ 
\midrule
QCML\cite{Ganscha2025} & 33.5M & PBE0-D4/NAOs\cite{Blum2009} 
     & Chemical graphs \\ 
 & & & Normal mode \\ 
 \midrule
AIMNet2\cite{Anstine2024} & 20M & $\omega$B97M-D3/def2-TZVPP 
     & ANI-2x\cite{Devereux2020}, OrbNet Denali\cite{Christensen2021} \\
 & & & Normal mode \\ 
 & & & Metadynamics \\ 
 & & & Molecular dynamics \\ 
 & & & Torsion scan \\
\midrule
$\nabla^2$DFT\cite{Khrabrov2024} & 16M & $\omega$B97X-D/def2-SVP 
     & Relaxation trajectory \\
\midrule
Transition-1x\cite{Schreiner2022} & 10M & $\omega$B97X/6-31G*
     & Nudged elastic band \\
\midrule
ANI-1x\cite{Smith2020}/ANI-2x\cite{Devereux2020} & 8.9M & $\omega$B97X/6-31G* 
     & Dimer and torsion scans \\
 & & & Normal mode \\
 & & & Molecular dynamics \\
 & & & Active learning \\
\midrule
QM7-X\cite{Hoja2021} & 4.2M & PBE0-MBD
     & Normal mode \\
\midrule
SPF\cite{Unke2019} & 2.7M & RPBE-D3(BJ)/def2-TZVP
     & Molecular dynamics \\
 & & & Active learning \\
\midrule
GEMS\cite{Unke2024} & 2.7M & PBE0-MBD/def2-TZVPP
     & SPF\cite{Unke2019} \\
 & & & Molecular dynamics \\
\midrule
SPICE\cite{Eastman2023}/SPICE2\cite{Eastman2024} & 2M & $\omega$B97M-D3(BJ)/def2-TZVPPD
     & Molecular dynamics \\
 & & & Dimer scan \\
\midrule
\midrule
\bf Materials Dataset & \bf Size & \bf Calculation method(s) & \bf Sampling method(s) \\
\midrule
OC20\cite{Chanussot2021} & 265M & RPBE/PAW 
     & Relaxation trajectory \\
 & & & Molecular dynamics \\
 & & & Rattling \\
\midrule
OMat24\cite{BarrosoLuque2024} & 110M & PBE(+U)/PAW 
     & Rattled Boltzmann \\
 & & & Molecular dynamics \\
 & & & Relaxation trajectory \\
\midrule
ODAC23\cite{Sriram2024} & 38M & PBE-D3(BJ)/PAW
     & Relaxation trajectory \\
\midrule
Alexandria\cite{Schmidt2024} & 30M & PBE(+U)/PAW, SCAN/PAW 
     & Relaxation trajectory \\
 & & PBEsol/PAW & \\
\midrule
OC22\cite{Tran2023} & 9.8M & PBE(+U)/PAW 
     & Relaxation trajectory \\
\midrule
MPtrj\cite{Deng2023} & 1.6M & PBE(+U)/PAW
     & Relaxation trajectory \\
\midrule
\midrule
\bf Unsupervised Dataset & \bf Size & \bf Calculation method(s) & \bf Sampling method(s) \\
\midrule
Zinc20\cite{Irwin2020}/Zinc22\cite{Tingle2023} & 4.5B & MMFF94 & Torsion sampling \\
\midrule
Uni-Mol\cite{Zhou2018}/Uni-Mol2\cite{Ji2024} & 838M & MMFF94 & Distance geometry \\
& & & Geometry optimization \\
\bottomrule
\bottomrule
\end{tabular}
\caption{Publicly available large scale datasets. We listed datasets of more than 1 million 3D structures with energy and force labels, as well as datasets of more than 100 million 3D structures without these labels.}
\label{tab:data}
\end{table}

\noindent
forces (>50 eV/\AA) can be detrimental to training and are usually filtered out, although high-energy and force data can be weighted with a Boltzmann factor\cite{Guan2023}, which still provides the model important chemical information about these regions. Finally, given that LLM FMs are pre-trained in an unsupervised manner on unlabeled data, it is worth asking what sort of data would be required to realize an analogous procedure in the chemical context. Ji and co-workers recently demonstrated that self-supervised pre-training on 800 million unlabeled 3D conformations of organic molecules improved down-stream performance on QM9 property prediction.\cite{Ji2024} 

\begin{figure}[H]
\centering
\includegraphics[width=0.99\linewidth]{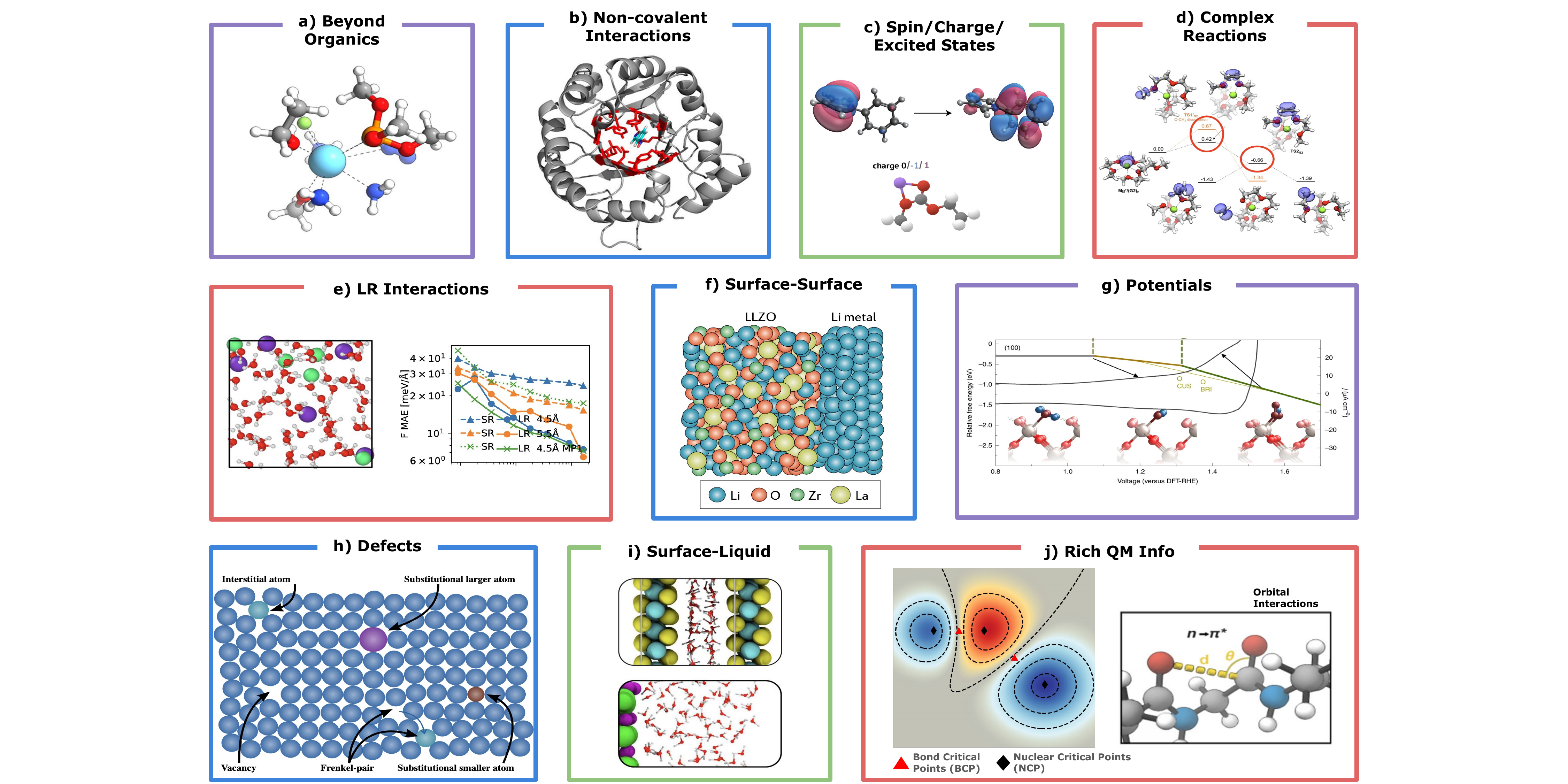}
\caption{Proposed areas where substantial additional high-quality DFT data should be generated in order to realize a general and transferable foundational MLIP. a) Beyond-organic elements. Reproduced with permission from Ref. \cite{Hlzer2024}. b) Non-covalent interactions.  Reproduced with permission from Ref. \cite{Vaissier2017}. c) variable charge/spin/excited states. Adapted with permission from Ref.\cite{liang2022revisiting,SpotteSmith2023}. 
d) Complex reactivity. Reproduced with permission from Ref.\cite{Seguin2019}. e) long-range interactions. Reproduced with permission from Ref. \cite{kim2024learningchargeslongrangeinteractions}. f) Solid-solid interfaces. Reproduced with permission from Ref. \cite{Xiao2019}. g) Surfaces under applied potential. Reproduced with permission from Ref. \cite{Rao2020}.  h) Defects. Reproduced with permission from Ref. \cite{Feri2011}. i) Solid-liquid interfaces. Reproduced with permission from Ref. \cite{mace-mp-0}. j) Rich quantum information. Left: Reproduced with permission from Ref. \cite{Vargas2024}. Right: Reproduced with permission from Ref. \cite{Boiko2024}.}
\label{fig:future_datasets} 
\end{figure}

The specific areas where we believe the community should devote major effort to large-scale data generation are shown in Figure \ref{fig:future_datasets}. For molecular DFT, the majority of the periodic table remains poorly covered, and under-representation of metal-organic complexes, complex electrolytes, metalloenzymes, and disordered proteins underscores the need for substantial dataset generation efforts beyond organic elements (Fig. \ref{fig:future_datasets}a). These types of systems could also address three other major gaps: non-covalent interactions (Fig. \ref{fig:future_datasets}b), systems where long-range interactions are critical (Fig. \ref{fig:future_datasets}g), and variable charge, spin, and excited  states\cite{liang2022revisiting, SpotteSmith2023} (Fig. \ref{fig:future_datasets}c) which have been poorly covered by current datasets. Excited state energies and properties also offer an avenue for dataset development, where time-dependent DFT (TDDFT) calculations\cite{dreuw2005single} are relatively inexpensive (roughly the cost of ground state DFT per state) and have been extensively benchmarked\cite{liang2022revisiting} against higher level theories\cite{loos2020quest}(Fig. \ref{fig:future_datasets}e).
We suggest that, beyond open-shell metals, combustion chemistry and atmospheric chemistry are spaces where different spin states are important and could be valuable. The final major gap we will describe is reactivity, particularly involving complex mechanisms and reactions in the condensed phase (Fig. \ref{fig:future_datasets}d). 

In the context of plane-wave DFT, we believe that major gaps include solid-solid interfaces (Fig. \ref{fig:future_datasets}h), solid-liquid interfaces (Fig. \ref{fig:future_datasets}j), surfaces under different applied potentials, (Fig. \ref{fig:future_datasets}g), and defects (Fig. \ref{fig:future_datasets}i). We note that some of these areas will require the use of better density functionals beyond the ubiquitous PBE and RPBE. Analogous to spin states for molecular properties, additional data generation for magnetic properties could also aid in correct magnetic moment initialization for diverse materials.

Finally, we believe there is a major need for large-scale datasets of rich quantum information, such as electron densities, orbital information, QTAIM descriptors, NBO interactions, and beyond (Fig. \ref{fig:future_datasets}k). We note that infusing rich quantum information into ML models has recently been shown to be beneficial\cite{Vargas2021,Vargas2024,Li2024_when,Boiko2024}, and believe that powerful FM capabilities could be unlocked with such data. For example, MLIP FMs could be pre-trained on this extra information by swapping prediction schemes to predict atom and bond-level values. This could pave the way for foundational chemical models that go beyond MLIPs, serving as powerful representation models that can be adapted to predict forces and energies downstream.

The primary point of a pre-trained MLIP FMs is that they can be fine-tuned to yield accurate PESs for diverse chemical systems, or to enable property predictions beyond just energies and forces. At present, DFT calculations provide the best trade-off between accuracy and computational cost, such that nearly all supervised large-scale datasets for MLIPs are comprised of DFT calculations. There are at least three categories of errors in the reference values of such datasets, any one of which alone could justify fine-tuning. First is the choice of functional (or wavefunction"al" if one goes beyond DFT). Standard GGAs yield RMS errors that are typically 2-4 times larger than leading hybrid functionals\cite{Mardirossian2017,Najibi:2018b} which in turn have RMS errors that are at least 10 times larger than "gold standard" coupled cluster theory\cite{helgaker2008quantitative} through perturbative triples, CCSD(T),\cite{Raghavachari:1989} or even beyond.\cite{karton2022quantum} This is a significant source of error. Turning from molecules to solids, there is an even greater need for high-quality data for fine-tuning. This begins with higher rung meta-GGA data in materials systems 
\cite{Smith2019,Kaser2020,Chen2023,Khazieva2024}. The second error source is the choice of basis set; at least triple and preferably quadruple-zeta is required to converge hybrid DFT molecular relative energies towards the complete basis set (CBS) limit\cite{Mardirossian2017} yet many of the large-scale DFT datasets summarized in Table \ref{tab:data} use only double-zeta basis sets (e.g. 6-31G*) that induce significant errors. In condensed matter, the corresponding choice of AO basis or plane-wave cutoff is equally important.\cite{witte2019push,bosoni2024verify}

Going beyond DFT is desirable. Here it is appropriate to highlight the effort of Smith and co-workers in generating the ANI-1ccx data set containing energies and forces of 500,000 organic molecules obtained with an accurate CCSD(T)*/CBS composite extrapolation.\cite{Smith2019,Smith2020}. This protocol has deviations from the Schrödinger limit that are larger than normally acceptable for small, very high accuracy datasets,\cite{karton2025good} but are much smaller than the best DFT errors, even in large basis sets.\cite{Mardirossian2017} Continued diverse collections of gold-standard CCSD(T) data (or beyond when necessary) would be extremely valuable for fine-tuning. Another result worth highlighting is the CCSD(T) assessment\cite{spiekermann2022high} of over 10,000 barrier heights in RDB7, which led to unexpectedly large RMS errors from the normally reliable $\omega$B97X-D3 functional. The largest of these deviations were resolved by removing orbital instabilities and allowing the DFT to be spin-polarized when necessary.\cite{Liu2025Revisit} However this highlights the limitations of even hybrid DFT for reactive systems where strong correlation effects are at play, as revealed by such symmetry-breaking.\cite{Shee:2021} For the most part, CC benchmarks for condensed matter periodic systems lie in the future, although there has been notable progress in CC methods for materials science.\cite{zhang2019coupled}

The third factor are other error sources that prevent a given ``model/basis'' combination from yielding a well-defined value. In particular, DFT calculations on materials require careful attention to best practices\cite{lejaeghere2016reproducibility,bosoni2024verify} to ensure reproducibility, beginning with choice of cutoffs, quadrature quality, and convergence criteria. Such factors have been discussed and investigated for the OC20 data.\cite{Abdelmaqsoud2024} There are also non-trivial additional sources of error. Reaching the thermodynamic limit is particularly challenging for hybrid functionals.\cite{quiton2024staggered} Pseudopotential errors, typically not present in molecular calculations involving light elements, also require careful evaluation\cite{borlido2020validation,rossomme2023good,Li2023}. Thresholds and cutoffs also affect molecular calculations, but can be controlled, and combined with SCF algorithms that ensure descent.\cite{VanVoorhis2002geometric} The local correlation methods needed for CCSD(T) on large molecules\cite{liakos2015possible} are also subject to domain errors.\cite{liakos2015exploring} They scale with size, and may impact accuracy\cite{santra2022performance,gray2024assessing} and so ideally should be removed and replaced by strict numerical thresholds.\cite{wang2023local,shi2024local}. These systematic errors make dataset fusion difficult as machine learning models trained on such datasets will likely have difficulty learning physical properties/labels from disjoint underlying distributions. In addition, reviewers could help find inconsistencies before initial publication and scientists remain committed to maintaining datasets after publication. One recent example underscoring these realities was a study by Garrison et. al.\cite{Garrison2023} where the authors trained machine learning models on the tmQM\cite{Balcells2020} dataset but found many structures in the dataset to contain missing hydrogen atoms. After filtering and recomputing the dataset, model performance improved significantly.

Looking beyond energies and forces, accurate spectral simulations and other response properties which intimately depend on 3D atomic positions would be extremely valuable for use in fine-tuning, dramatically expanding the properties that fine-tuned FMs could predict. Finally, fine-tuning atomistic FMs with experimental data such as radial distribution functions, solvation free energies, density, and viscosity, could also be desirable. We note that experimental data can also be used in the context of selecting or "aligning" FMs trained on small cluster datasets to properties better reflected by bulk simulations. For example, Gong et al. map MLIP forces to pressures (and thus densities) from a small set of representative, experimental measurements\cite{Gong2024}. We discuss additional training strategies that incorporate experimental data in Section \ref{subsec:train}. 

\subsection{Model-Agnostic Training Strategies}
\label{subsec:train}
\noindent
Effective training strategies have and will continue to play an important role in developing the next generation of FMs for chemistry and materials. We highlight multiple directions relevant to both pre-training as well as post-training schemes, with the goal of obtaining ``model-agnostic'' training strategies that can be applied to any foundational MLIP.

\textit{Pre-training strategies.} While unsupervised training methods have been critical in improving the foundational capacity of LLMs and LVMs, in computational chemistry and materials science there are typically energy and/or forces labels (or molecular property labels for Zinc22\cite{Tingle2023}/Zinc20\cite{Irwin2020}) associated with atomistic configurations such that supervised learning is the common approach for MLIP optimization. By contrast self-supervised approaches start with the key ingredient being the availability of a data distribution that might sensibly center on molecular geometry, for instance, the PCQM4Mv2\cite{Hu2021} and PM6\cite{Nakata2020} datasets developed under the PubChemQC Project\cite{Nakata2017}. This type of data is highly suitable for graph self-supervised learning using denoising that enables pre-training models to learn from molecular graphs for intrinsic chemical information. One recent example used geometric denoising to create a pre-trained model that exhibited fine-tuned accuracy improvement compared to randomly initialized "from scratch" models for various chemical properties in the QM9 dataset.\cite{Zaidi2022,Neumann2024} This denoising approach was later extended to work for non-equilibrium structures \cite{liao2024generalizingdenoisingnonequilibriumstructures}. Understanding how much self-supervised pre-training relies on the quality of the training distribution, and how scalable such approaches can be in terms of data size compared to purely supervised (pre-)training, is currently an active area of investigation. An additional approach that is gaining acceptance is pre-training using synthetic energy and force data provided by other from-scratch trained MLIPs, which has been demonstrated to improve robustness of fine-tuned NN potentials on the condition that the synthetic source is reasonably reliable.\cite{Gardner2024} Other examples of self-supervised learning approaches in materials chemistry include twin algorithms\cite{Magar2022} and atom replacement\cite{Sakai2023} pre-training schemes.

Although most current models train on a dataset with calculations done at one level of theory \cite{Batatia2023,Kovacs2023}, future foundational MLIPs could benefit from training on multiple levels of theory. A training objective that extracts learning signal from multiple levels of theory could take advantage of more abundant datasets generated with cheaper methods (like a semi-empirical potential), while still maintaining accuracy by leveraging higher fidelity data. Incorporating different levels of theory into the training pipeline would also allow MLIPs to learn about different interactions that are better described by different methods \cite{Shoghi2023,Pasini2024,Zhang2024}. Having multiple types of datasets also presents challenges in training a model that can handle different energy scales and the nuances of each level of theory. However, addressing these challenges could lead to the development of more general MLIPs.

Physical laws can also be incorporated into MLIPs by placing it directly into the loss function and training procedure, instead of making them inherent to the model architecture. For instance, rotational equivariance can be enforced via data augmentation loss as discussed previously, and energy conservation can be encouraged by minimizing the curl of the MLIP as part of the training objective\cite{Chmiela2017}. While these ``soft'' constraints do not guarantee that the law will be obeyed, they circumvent the need to design new architectures for each new constraint of interest, and they can enable more scalable and flexible training\cite{Pan2021}.

\textit{Post-training strategies.} After pre-training a FM, post-training is an important procedure to further refine and optimize the model to a given downstream application. For domain specific tasks, such as energy and force prediction, modest amounts of higher-quality and labeled data are used within a standard supervised fine-tuning (SFT) context. Model distillation is another post-training approach for fine-tuning. In model distillation, a simpler model (the "student") is trained to replicate the behavior and performance of a larger, more complex model (the "teacher", or FM in this case). This process involves transferring knowledge from the teacher to the student by using the teacher's predictions or internal representations as targets for training the student model. The use of teacher-student constructs frequently allows to perform more computationally efficient training, particularly for multi-label training cases.

\begin{figure}[H]
 \centering
 \includegraphics[width=0.95\textwidth]{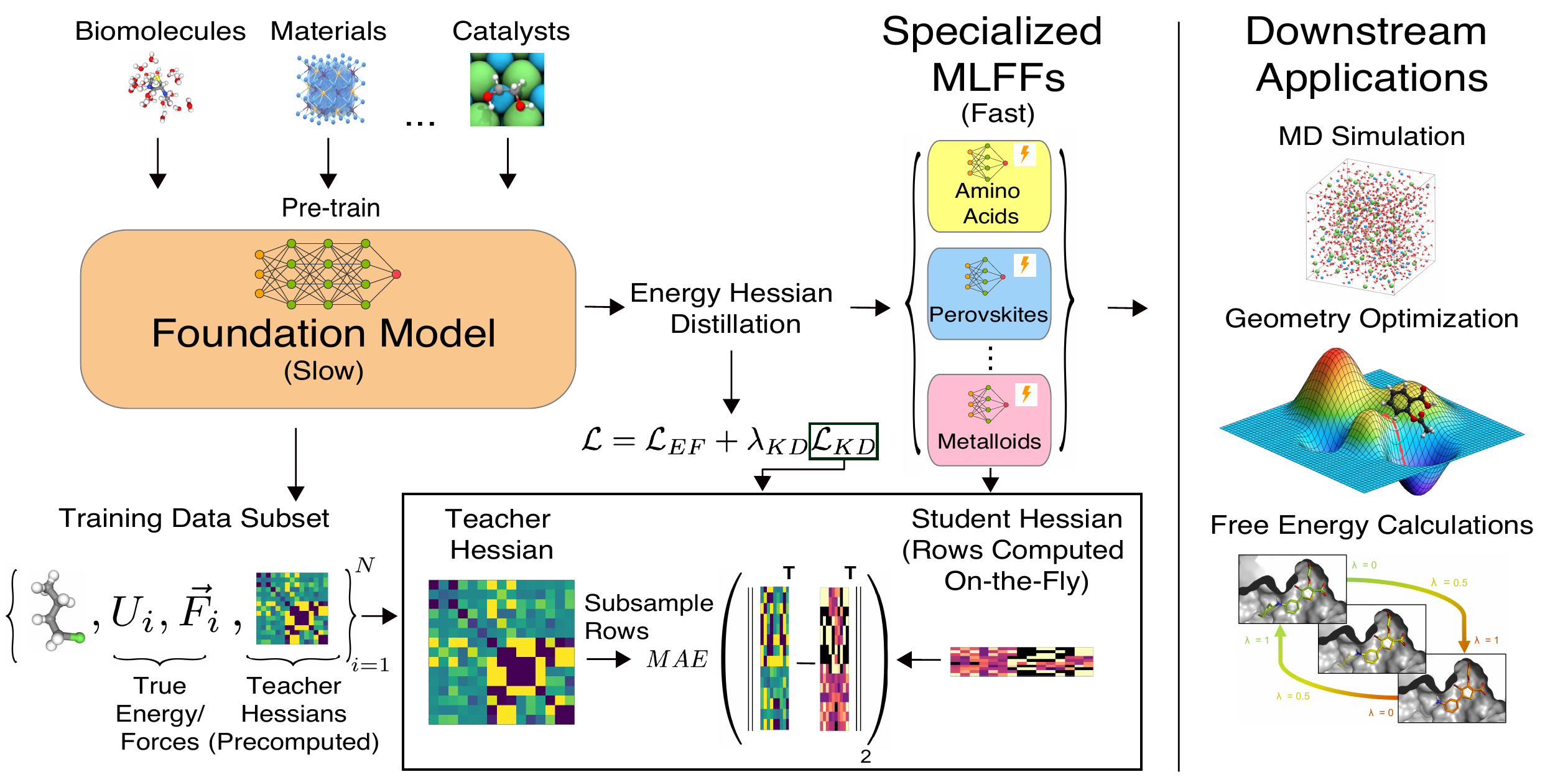}
 \caption{\textit{Example of a post-training strategy for machine learning interatomic potentials via model distillation.} One starts with an MLIP FM that has been trained on a large quantity of diverse data. A series of smaller MLIPs can then be trained via a knowledge distillation procedure, enabling specialization and speed for a specific task while retaining the general-purpose representations learned by the large-scale FM. Reproduced with permission from Ref.~\cite{Amin2025}.
 }
 \label{fig:model-distillation}
\end{figure}

Kelvinius et al. performed MLIP distillation by aligning node and edge features across models, but did not consider the task of specializing from general-purpose FMs.\cite{Kelvinius2023} Recently Amin and co-workers\cite{Amin2025} used distillation  on large-scale models like MACE-OFF \cite{Kovacs2023}, MACE-MP-0 \cite{Batatia2023}, and JMP \cite{Shoghi2023}. By matching the Hessians of the energy predictions, these large-scale models were distilled into small, student MLIPs specialized for a specific chemical subset of the data and up to 50 $\times$ faster during inference-time compared with the original large-scale model (see Fig. \ref{fig:model-distillation}). This approach is agnostic to the MLIP architecture, and is compatible with arbitrary combinations of teacher and student MLIP design choices. Model distillation is one potential solution to balance scalable training with ensuring physical consistency in the final simulation. For instance,\cite{Amin2025} showed that a large-scale model can be scalably trained without expensive constraints such as gradient-based forces for energy conservation, and can then be distilled into a student MLIP possessing gradient-based forces. This ensures energy conservation while leveraging the general-purpose representations from the pre-trained large-scale model.

Thus far, much of the discussion has been about MLIPs trained on ab-initio data. However, experimental data is also a rich potential source of information to train MLIPs. Directly incorporating experimental data into MLIP training has primarily been done through differentiable simulation\cite{Wang2020,Thaler2021,sipka2023,Navarro2023,raja2025stabilityaware,Gangan2024}. In this case, MLIPs are typically initialized by pre-training with an analytical prior potential, or on a sparse set of simulated observables such as radial distribution functions, phonon densities, and diffusivity coefficients, which can be computed as ensemble averages over an equilibrium MD simulation. The pre-trained model is then fine-tuned to match experimental observables. Computing gradients necessary for MLIP finetuning through the full MD simulation is prohibitively expensive, and so techniques such as reweighting \cite{Thaler2021}, implicit differentiation \cite{Blondel2022EfficientImplicitDiff}, and the adjoint method\cite{chen2019neural} are leveraged to enable tractable optimization. However, major technical challenges remain, the most notable of which is effectively training with \textit{dynamical} observables \cite{Bolhuis2023OptimizingMolecularPotentials}. 

\subsection{Criteria for a Successful Foundation Model}
\label{subsec:eval}
\noindent
A MLIP FM for atomistic simulation should perform well on a diverse range of tasks after fine-tuning and post-training, spanning different levels of theory, different types of systems, static and dynamic quantities, and a range of properties beyond just energies and forces. Unrelated downstream tasks might include field-dependent property predictions, generative models for crafting new structures or molecules\cite{Zaidi2022,Magar2022,Wang2023,Gardner2024,Yang2024}, measuring energies and forces in defective structures\cite{Zuo2020, Liu2023}, and dynamic testing environments created through MD simulations under more varied conditions.\cite{Fu2022,Kapil2022,Liu2023,Guan2023} 

One critical goal is the development of testing platforms for MLIP FMs to test their endowed capabilities. It demands considerable effort with regard to the computational cost of many simulation runs, studies to validate testing components, and analyses to derive scientific insights. But we believe that the benefits to the scientific community are profound and represent an essential direction for future research in demonstrating the promise of MLIP FMs.\cite{Fu2022,Wang2022,Rohskopf2023,Liu2024} While competition over machine learning model performance on small datasets like MD17\cite{Chmiela2017} did drive crucial early MLIP development, a number of large community efforts have emerged more recently that define a successive series of "leaderboards" as testing grounds for MLIPs. The OC20 leaderboard\cite{Chanussot2021} was established in 2021, with a dataset split into training/validation/testing categories with rigorously defined in-distribution (ID) and out-of-distribution (OOD) systems, and a private test set to prevent test performance from being gamed. Energy and force model performance among methods developers on the OC20 leaderboard has consistently improved over time since its inception. In 2023, the MatBench Discovery leaderboard\cite{Riebesell2023} was created to test MLIP performance for crystalline materials relaxations and the ability to predict materials thermodynamic stability and formation energy. While initially focused on the use of only MPtrj\cite{Deng2023} as the training dataset, 2024 saw an explosion of interest in MatBench Discovery, including teams from Google, Microsoft, and Meta, each reporting the generation of supplemental training data as well as models trained on their data beating the previous best performing model. However, some of the new data may risk leakage with the test set. Further, most of the leaderboard evaluations are overly focused on energy, which may be overfit in order to improve leaderboard ranking while degrading actual scientific utility. 

A related issue is the lack of robust uncertainty quantification. The prevalent method for quantifying uncertainty in MLIP predictions involves using ensembles \cite{Lakshminarayanan2016,Peterson2017,Imbalzano2021}; this approach entails running multiple fits of the MLIP simultaneously and interpreting the variances among them as a measure of uncertainty. However, ensembles for quantifying uncertainty is computationally demanding, requiring parallelization, and may be unsuitable for MLIP FMs of larger scale. Such prediction uncertainty has not thus far been taken into account in previous leaderboards. Both issues have severe consequences in practical applications, and proper tests beyond simple accuracy measures have to be carefully designed.

Even if the MLIP universality within a specific domain can be evaluated through OOD tests, the leaderboards do not seek to show that the resulting models are foundational, i.e. that they obey scaling laws or are derived from diverse data or pre-training strategies or that they exhibit superior performance on a broad range of down-stream tasks after fine-tuning. While the large size of the OC20 training set makes it a viable arena in which to evaluate model scalability, e.g. via performance with ~1M vs ~10M vs ~100M training data, only performance on the full training dataset is featured on the leaderboard. Further, both OC20 and MatBench Discovery focus on well-behaved relaxations but without consideration of other emergent properties such as MD stability or PES smoothness, which has been shown to be independent of just test errors for energies and forces.\cite{Fu2022,Morrow2023,Wang2023,Bihani2024} However the recently released NNP Arena\cite{nnparena} is a step in the right direction, by evaluating energy error for both pre-trained MLIPs (and pure DFT models) against gold-standard coupled cluster reference values, without any concern for what data was used in MLIP training or how MLIP training was performed. NNP Arena further evaluates inference speed, which previous leaderboards haven't considered but which may be a very important consideration for the future. Another important direction pursued in the recent MLIP Arena\cite{chiang2025mlip} involves assessing the extent to which models comply with fundamental physics, including energy conservation, asymptotic behavior, smoothness, symmetry, and simulation stability. These metrics do not require additional labels, yet they carry substantial implications for the reliability and applicability of the models in practical settings.

\subsection{Toward Foundation Models in Chemistry and Materials}
\label{subsec:eval}
\noindent
At present all MLIPs are not true FMs by all criteria outlined in Section 2, but there are a few prototype MLIP FMs on the horizon that are proving more accurate, to more comprehensively cover the periodic table for molecular systems, and also show the ability to comply with heuristic scaling laws. In particular the pretraining models such as Efficiently Scaled Attention Interatomic Potential (EScAIP)\cite{Qu2024} and Uni-Mol2\cite{Ji2024} have characterized the scaling correlations of decreasing validation error with increasing model size, dataset size, and computational resources. This would suggest that emergent capabilities might follow, such as predicting energy labels from unlabeled data as shown for Uni-Mol2\cite{Ji2024}. The fact that LLMs have demonstrated such scaling laws has led to emergent properties in chemistry for generative tasks in small molecule drugs\cite{Cavanagh2024} and their synthesis\cite{Sun2025}, although the LLMs were not trained on SMILES strings or chemical reaction templates.

Since the initial writing of this manuscript, two important contributions have been released which substantially advance the state-of-the-art for atomistic simulation: the Open Molecules 2025 (OMol25) dataset\cite{levine2025}, and the Universal Model of Atoms pre-trained MLIP\cite{UMA2025}. OMol25 is the largest ever molecular DFT dataset for training MLIPs, with over 100 million snapshots at an excellent level of DFT ($\omega$B97M-V/def2-TZVPD), where snapshots include up to 350 atoms, up to ten unpaired electrons, total charges ranging from -10e to +10e, and 83 elements of the periodic table. OMol25 broadly covers small molecules, electrolytes, biomolecules, and metal complexes, using more than ten distinct methods for structural sampling.\cite{levine2025} OMol25 thus makes substantial progress towards high-quality data coverage beyond organics, such as non-covalent interactions, spin and charge species, complex reactions, and long-range interactions, nearly half of the areas outlined in Figure \ref{fig:future_datasets}. 
In addition to the $>$100 million training data, OMol25 also includes explicit out-of-distribution test sets and novel model evaluations, which will populate a public leaderboard in the near future, and baseline trained MLIPs.

Simultaneous with OMol25's release was the release of UMA\cite{UMA2025}, which is trained on OMol25\cite{levine2025}, OMat24\cite{BarrosoLuque2024}, ODAC23\cite{Sriram2024}, OC20\cite{Abdelmaqsoud2024}, and the novel Open Molecular Crystal dataset (unpublished), totaling nearly 500 million training data - the largest and most diverse training set ever used to train a single MLIP. UMA also demonstrates scaling laws for both number of learnable parameters and compute, thus coming the closest to our vision of a foundation model for atomistic simulation. UMA is based on the eSEN architecture\cite{fu2025} and further employs a novel mixture of linear experts (MoLE) for two key benefits: 1) the MoLE causes a model trained on all datsets simultaneously, in a multi-task fashion (where each distinct training dataset is a distinct "task"), to exhibit superior performance to models trained separately on each dataset. 2) the MoLE allows a huge number of learnable parameters to be active during training, but a much smaller number to be active at inference time, allowing the model to effectively leverage the vast training dataset while still remaining fairly fast and memory efficient for subsequent simulations.

UMA and eSEN-OMol now top Rowan's NNP Arena leaderboard\cite{nnparena} for neutral, closed-shell, organic molecular energies, with errors against coupled cluster for the GMTKN55 benchmark set falling by nearly 5 kcal/mol compared with the previous best model. Digging more deeply into the GMTKN55 test set\cite{Goerigk2017}, one can see that UMA's energies are already lower error with respect to CCSD(T) than $\omega$B97M-D3BJ/def2-QZVP DFT for variable charge, variable spin, and metal-containing large systems and barrier heights, while being nearly comparable for intermolecular non-covalent interactions. However, the OMol25 evaluation tasks demonstrate that there is still substantial room for improvement on charge, spin, and long-range interactions, where error against reference DFT data remains very large ($>100$ meV) likely due to the naive handling of charge and spin and the 6\AA\ graph cutoff, emphasizing the key areas where further MLIP architecture development is needed.

\section{Discussion and Future Directions}
\noindent
The primary promise of a MLIP foundation model is that it would enable massively impactful investigation of systems ranging from metal-organic frameworks to enzymes to electrocatalysis beyond current physics-only models, while avoiding the need to build each ML model from the ground-up for any new chemical system of interest. One key insight that has been observed in other areas of machine learning is that methods that may work in ``low data'' regimes do not necessarily work as well in higher data regimes. As data size increases, as we have summarized for the many large-scale community and data efforts in chemistry and materials, can ML architectures scale accordingly to handle such data? Currently, the two paradigms are to use inductive biases to address physical constraints, often requiring less data\cite{Batzner2021,Haghighatlari2022}, with an alternative philosophy being that such physical laws can be easily learned when there is enough data, and when model capacity and compute is correspondingly scaled to take advantage of this data. Constrained MLIPs or learning energy and forces together ensures that conservative forces are available for the primary downstream MD task\cite{Pan2021}, but post-training strategies can take non-conservative MLIPs and distill them into conservative MLIPs\cite{Amin2025}, enabling both initial training at scale and physical soundness at deployment. Hence we posit here that incorporation of architectural inductive biases is not the only way to create sound MLIP FMs, a direction that is currently an underexplored paradigm in chemistry and materials. It raises the interesting question whether the chemical and materials sciences will escape the ``bitter lesson'', or whether our science domains may also conclude that data and scalability is superior to built-in constraints, as has come to pass in other fields such as robotics and weather that are also grounded in physics. 

Training a truly foundational MLIP that can realize the far-reaching impact we envision will require a collection of very large datasets that broadly span chemistry, materials, phases, and interfaces. While our earlier discussion focused on using DFT-labeled data as the standard approach for pre-training, perhaps even much larger low-level datasets with or without labels from semi-empirical methods or classical force fields or on-the-fly schemes for computing properties could be important to unlock powerful at-scale capabilities for chemical FMs. It is also possible that the best approach will not actually start with an MLIP, but will instead leverage some combination of traditional chemoinformatics, chemically-relevant text, and/or un-labeled structures to create datasets suited for larger FMs which can be retooled as MLIPs. Information contained in other multi-modal data that incorporates experimental information could also be extremely valuable for pre-training, although realizing this in practice will require massive data curation efforts in order for the scale of the data to be suitable for pre-training as well as the development of novel training strategies even beyond what we have discussed to be viable at scale. 

While current MLIP testing platforms do provide a substantial resource to the community, there are still significant challenges and opportunities in the context of MLIP FMs. For example, there is not always a direct correlation between reduced force errors observed in test set performance and its actual performance in MD simulations.\cite{Fu2022,Morrow2023,Wang2023,Bihani2024}. Current MLIP leaderboards also do not yet explicitly evaluate model scalability with respect to the amount of training data or learnable parameters, nor do they seek to quantify the "foundational capacity" of the model via standardized fine-tuning across chemical and materials domains and properties. Scaling laws being the defining property of a FM, MLIPs should demonstrate continuous improvement with more parameters, more data, and more compute. Ultimately the FM practitioner must demonstrate the success of a FM via fine-tuned models that are fast and specialized - and which are more accurate than a model trained from scratch on data specific to that task/domain. The development of a scalable FM will also require using efficient algorithms to perform pre-training and fine-tuning. In particular, efficient mechanism for storing and retrieving data for batched training, dynamic load balancing methods that can accommodate variations in graph sizes, and communication hiding techniques that overlap computation and message passing will be important for achieving a rapid turnaround.

We note that the success of foundational LLMs and LVMs has critically depended on acquisition of vast collections of text and images from the internet, i.e. data obtained via implied or enforced open access. But in the context of DFT data for training MLIPs, this assumption of open access to data is not always adhered to when commercial players are involved. For instance, the datasets used to train Google's GNoME model for crystal energies and Microsoft Research reported MatterSim have not been released for public use. In contrast, the FAIR Chemistry team at Meta, responsible for the largest MLIP training datasets built to-date, has consistently released their datasets (and trained models) fully open-source, a value to the scientific community that cannot be understated. In a related vein, MLIP models have benefited tremendously from industry-released software tools such as TensorFlow\cite{tensorflow2015-whitepaper}, Keras\cite{chollet2015keras}, and PyTorch\cite{paszke2017automatic}. However, open-access development has not occurred for LLMs such as ChatGPT\cite{OpenAI_2023}, and the loss of implementation details on how LLM FMs work could limit our ability to develop similarly powerful MLIP FMs. We note that while DeepSeek-V3 is open source, and can deliver high-quality results at a lower cost compared to close-source LLMs, it is either restricted or outright banned in several countries due to concerns about ethics, privacy, and security. As is needed for all emerging technologies, training in ethical development and use of AI/ML should remain a top priority in the physical sciences.\cite{thg2024} An additional practical challenge is the computational resources required for the demonstration of FM capabilities at large scale. It will require world-class high performance computing beyond the academic lab scale and the first place to turn to are the leadership computing levels at the national labs. The combined issues of open source data, software, models, politics, and ethics training will likely require an even larger federation of government agencies with industry, and international cooperation, modeled on the paradigm of scale of the Human Genome Project\cite{Cook-Deegan1994}.

\bibliography{references}
\bibliographystyle{naturemag}

\section{ACKNOWLEDGMENTS}
\noindent
The authors thank Eric Qu, Yingze Wang, Joseph Cavanagh, Kunyang O. Sun, Kareem Hegazy, Joseph P. Heindel, Peichen Zhong, Anup Kumar, Senwei Liang, Selim Sami, R. Allen LaCour, Sruthy K. Chandy, Ruoqi Zhao, and Dorian Bagni for their contributions in discussions on August 6, 2024 at UC Berkeley. T.H.-G. thanks the CPIMS program, Office of Science, Office of Basic Energy Sciences, Chemical Sciences Division of the U.S. Department of Energy under Contract DE-AC02-05CH11231 for support of the machine learning research. S.M.B. acknowledges support by the Center for High Precision Patterning Science (CHiPPS), an Energy Frontier Research Center funded by the U.S. Department of Energy, Office of Science, Basic Energy Sciences at Lawrence Berkeley National Laboratory under contract DE-AC02-05CH11231. S.V. was supported by the DOE's National Nuclear Security Administration's Office of Defense Nuclear Nonproliferation Research and Development (NA-22) as part of the NextGen Nonproliferation Leadership Development Program. This work was also supported by the U.S. Department of Energy, Office of Science, Office of Advanced Scientific Computing, and Office of Basic Energy Sciences, via the Scientific Discovery through Advanced Computing (SciDAC) program for the mathematical and computing foundations of machine learning models. This work was also supported by the U.S. Department of Energy, Office of Science, Energy Earthshot initiatives as part of the Center for Ionomer-based Water Electrolysis at Lawrence Berkeley National Laboratory under Award
Number DE-AC02-05CH11231, as well as Toyota Research Institute as part of the Synthesis Advanced Research Challenge. This work used computational resources provided
by the National Energy Research Scientific Computing Center (NERSC), a U.S. Department of Energy Office of Science User Facility operated under Contract DE-AC02-05CH11231.

\section{AUTHOR CONTRIBUTIONS STATEMENT}
\noindent
S.M.B., B.C., A.K., and T.H.-G. defined the topics and scope of the perspective. E.C.-Y.Y., Y.L., J.C., S.R., T.K., S.V., W.X. made the figures. E.C.-Y.Y., Y.L., J.C., S.R., T.K., S.V., P.Z., W.X., M.H-G., S.M.B., B.C., A.K., and T.H.-G. wrote the paper. All authors discussed the results and made comments and edits to the manuscript.

\section{COMPETING INTERESTS STATEMENT}
\noindent
The authors declare no competing interests.

\section{Graphical abstract}. 

\begin{figure}[H]
 \centering
 \includegraphics[width=0.5\textwidth]{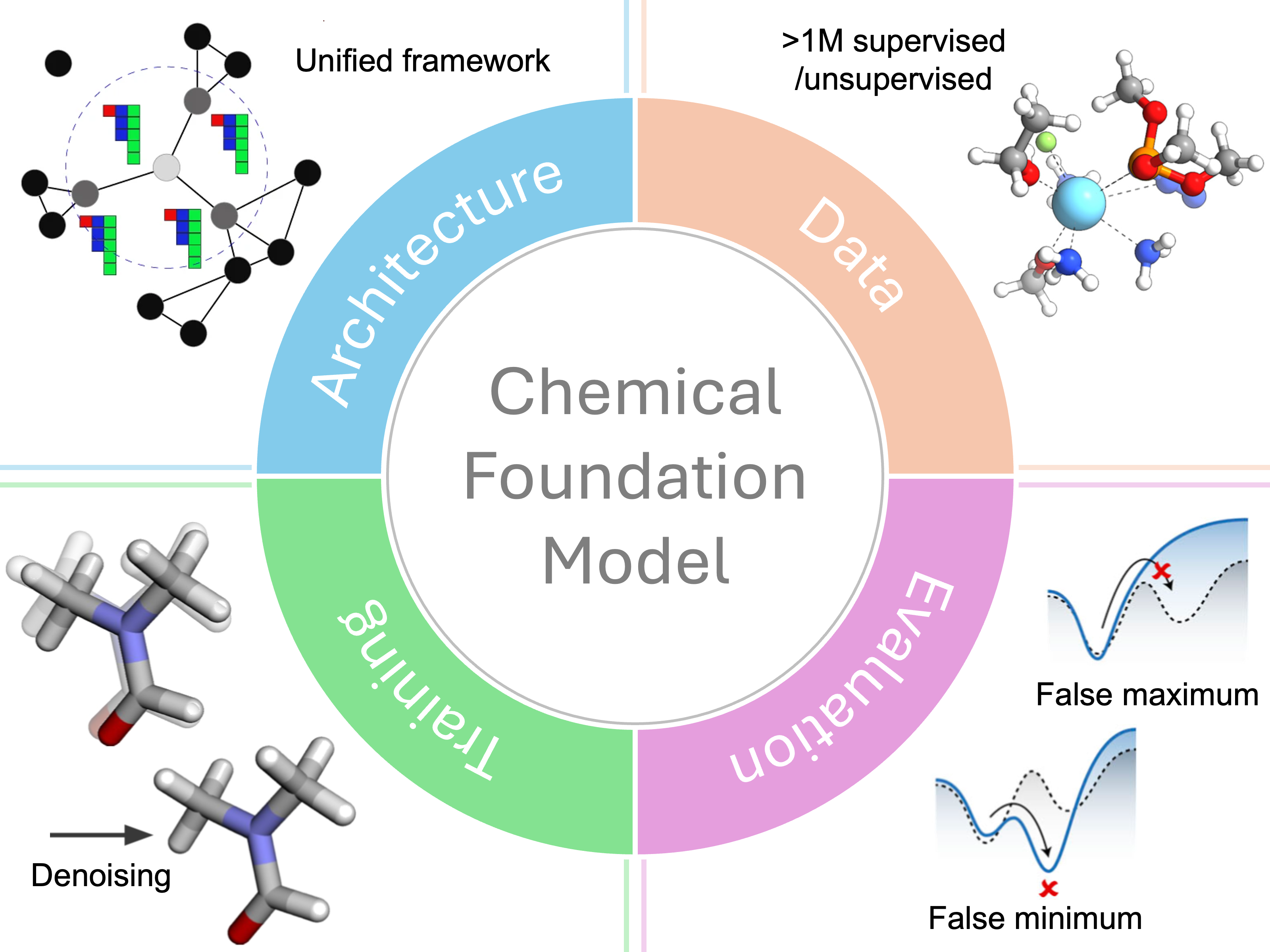}
 \label{fig:toc}
\end{figure}

\section{Short Summary}
\noindent
We examine the historical development and underlying principles of foundation models realized in language and vision, and propose how physics-infused machine learning interaction potentials could dramatically transform at scale to create transformative foundation models for chemistry and materials science. 



\end{document}